\documentclass[12pt]{article}
\usepackage[xdvi]{graphicx}
\usepackage{amssymb}
\newcommand{\bea}{\begin{eqnarray}}
\newcommand{\eea}{\end{eqnarray}}

\makeatletter
\@addtoreset{equation}{section}
\makeatother

\newcommand{\siml}{%
\hspace{0.3em}\raisebox{0.4ex}{$<$}\hspace{-0.75em}\raisebox{-.7ex}{$\sim$}\hspace{0.3em}}

\begin{document}

\begin{titlepage}
\begin{flushright}
OU-HET 655/2010
\end{flushright}

\vspace{25ex}

\begin{center}
{\Large\bf 
Low energy kinetic distribution on orbifolds
}
\end{center}

\vspace{1ex}

\begin{center}
{\large
Nobuhiro Uekusa
}
\end{center}
\begin{center}
{\it Department of Physics, 
Osaka University \\
Toyonaka, Osaka 560-0043
Japan} \\
\textit{E-mail}: 
uekusa@het.phys.sci.osaka-u.ac.jp
\end{center}


\vspace{3ex}

\begin{abstract}

Fermion self-energy associated with wave
function renormalization is studied
in a five-dimensional Yukawa theory
on the orbifold $S^1/Z_2$.
One-loop divergence can be subtracted
with only two renormalization constants
in the bulk and on the branes.
We show that the bulk and brane parts of the
self-energy are uniquely determined with
requiring physical conditions.
With this procedure,
momentum-scale dependence of the 
renormalized self-energy
is given and the
distribution of the bulk and brane parts
can be found at low and high energies.
Despite possible higher degrees of 
divergence in higher dimensions,
the regularization scheme dependence 
does not arise.
A viewpoint of the regularization scheme
dependence
at higher-loop level is also discussed.
We find that
the ratio of the bulk
contribution to the brane contribution
depends on the momentum scale in a very
mild way, 
so that
the relative coefficient of 
bulk and brane kinetic terms
can be regarded as 
approximately constant for the leading
quantum effect.
The physical conditions given here
are applicable to
remove ambiguity 
in various orbifold models.

\end{abstract}
\end{titlepage}

\newpage

\section{Introduction}

The picture that physical quantities depend
on the energy scale of interest is intuitive and
has provided a clue to understand laws of Nature.
A probe with a shorter wavelength
in a system would resolve
its substructure in a more microscopic way.
If the substructures in the system
are hierarchical,
it can achieve predictability
without knowledge of details at the other energy 
scales of the system.
In a renormalizable 
quantum
field theory in four dimensions,
information for short and long distances is
described well.
Divergence that the theory can have for short distance 
is subtracted.
After renormalization conditions are imposed,
physical quantities are finite
and they are energy-scale dependent.

The energy-scale dependence of physical 
quantities 
is important not only for 
renormalizable interactions but
also for non-renormalizable interactions.
Renormalizable interactions are treated
with no new counterterms
once all the renormalizable interactions are included.
In usual four-dimensional models,
non-renormalizable interactions are supposed to be
suppressed by the ratio of the energy scale
of interest
to the ultraviolet momentum cutoff of the theory.
In a four-dimensional theory where renormalizable and
non-renormalizable interactions 
coexist, physical quantities can be 
dominated by
contributions from
lower-dimensional operators being 
renormalizable terms.
In a theory with compactified extra dimensions,
fields as four-dimensional modes can have
dimension-four operators which are similar
to renormalizable terms in four dimensions.
From a standpoint that non-renormalizable 
interactions are irrelevant operators
and that their contributions to physical quantities
are negligible,
it is interesting to search for
rules or orders for possible effects
in higher-dimensional field theory
at each given loop level.

In a theory with compactified extra dimensions,
several characteristic properties
need to be taken into account.
They are higher-dimensional operators,
regularization scheme dependence and
brane terms.
Even if the starting action integral includes
all the operators up to a certain mass dimension,
radiative corrections give rise to
divergence for new local operators.
Because higher-dimensional operators needed
in the starting action integral
affect the values of
physical quantities,
they must be determined in some ultraviolet
completion or they must be found to be very small.
As for scheme dependence,
the point is that
degrees of divergence in higher dimensions can be 
higher.
For example, if
a four-dimensional integral produces logarithmic divergence,
the corresponding five-dimensional divergence
is expected to be linear divergence for
a naive cutoff regularization.
This linear divergence can be 
missed in a dimensional regularization.
It needs to be examined
whether such a regularization dependence occurs
for physical effects.
Finally for brane terms, it was found that 
loop effects of bulk fields
produce infinite contributions to require
renormalization by 
couplings on branes~\cite{Georgi:2000ks}.
Brane terms can be mass and kinetic energy
terms and higher derivative operators can be
needed as counterterms for loop corrections~\cite{%
vonGersdorff:2002as}-\cite{Muck:2004br}.
Research on effects of brane terms on 
mode functions and mass spectrum
has also been developed in the literature~\cite{%
Dvali:2001gm}-\cite{Ghilencea:2006qm}.
In principle,
the coefficients of brane terms and 
higher-dimensional operators can be 
energy-scale dependent.
In addition to divergence,
the finite part of quantum corrections
needs to be extracted for 
physical quantities.

In examining quantum loop effects,
two-point functions include
nontrivial information in higher-dimensional
field theory.
In $\phi^4$ theory in flat five dimensions
on the $S^1/Z_2$,
at one-loop level, there is no wave function renormalization.
At two-loop level, divergences 
for wave function renormalization appear 
not only for $p^2$ 
but also for $(p^2)^2$.
This higher-derivative term gives important information
for consistency of the theory.
The predictability of the theory with $(p^2)^2$
term
requires
the ultraviolet cutoff orders of magnitude
larger compared to the compactification scale~\cite{%
Uekusa:2009dy}.
One-loop wave function renormalization
appears in Yukawa theory.
On an orbifold, fermion self-energy has
divergence on
the branes~\cite{Georgi:2000ks}.
The corresponding fermion kinetic term
not only in the bulk but also on the branes
need to be included in the starting action integral.
At first sight,
the relative coefficient for
the bulk and brane kinetic terms
seems arbitrary.
It is the case at tree level of the action integral.
However, for the energy-dependence of physical
effects,
it is crucial to include
quantum effects.
When one extracts such effects,
it must be treated carefully whether the relative
coefficient is
only apparent ambiguity.
An analogy lies in 
four-dimensional renormalizable field theory
where coupling constants in tree-level action integrals 
are free parameters. After renormalization, 
they are physical and momentum dependent. 
The difference is that 
in higher-dimensional field theory
full information of non-renormalizable interactions
is unknown. 
To proceed phenomenological study without such an
information,  
one might be content to assume that
the relative coefficient can have various values.
In this sense, the coefficient is akin to
a free parameter.
One important point is to pay attention to 
physical consequences.
It is nontrivial whether all the non-renormalizable
 interactions are necessary for
making a physical prediction at
 low energies. 
Then one needs the relative coefficient that is not
a free parameter and its behavior that is 
found at low energies,
whereas
the momentum-scale dependence of the 
relative coefficient
has not been examined even in the leading level
in the literature.

In this paper, we examine
the momentum-scale dependence of fermion self-energy
associated with wave
function renormalization
in Yukawa theory
in flat five dimensions on the $S^1/Z_2$.
At one-loop level, wave function renormalization 
is needed as in four dimensions.
One-loop divergence can be subtracted
with only two renormalization constants
in the bulk and on the branes.
We show that the bulk and brane parts of the
self-energies are uniquely determined with
requiring physical conditions.
Our physical condition is that
fields with their physical masses obey 
usual Feynman propagators.
This is sufficient because
a relative difference between
left- and right-handed components
for fermions is 
related to 
the difference between the coefficients
of the bulk and brane kinetic terms.
With this procedure,
the momentum-scale dependence of the 
renormalized self-energies
is given and the
distribution of the bulk and brane parts
can be found at low and high energies.
While
the loop-momentum integral depends on
the regularization scheme,
the renormalized self-energy
is scheme-independent.
We find that
the ratio of the bulk
contribution to the brane contribution
depends on the momentum-scale in a very
mild way,  
so that
the relative coefficient of 
bulk and brane kinetic terms
can be regarded as 
approximately constant for the leading
quantum effect.

The physical condition given here
is applicable to
remove ambiguity 
in various orbifold models.
Following the recent discovery of 
marginal and interacting operators in models
with extra dimensions~\cite{ue},
we also give the low-energy behavior of
bulk and brane kinetic terms by the distance-rescaling 
for integrating 
out the shell of
high-momentum degrees of freedom.
It would be important to speculate
what arises at higher-loop level.
The regularization-scheme independence may not be
kept beyond the leading level.
We discuss a viewpoint of 
the regularization-scheme dependence
for physical quantities at higher-loop level.

The paper is organized as follows.
In Section~\ref{sec:model}, the model with 
brane kinetic terms is given.
The basic idea of our proposal 
to determine the relative coefficient
is described.
In Section~\ref{sec:subtract},
the one-loop divergence is given 
and is subtracted with the corresponding
counterterms.
In Section~\ref{sec:momentum},
the physical condition is given for renormalization.
The momentum dependence of fermion self-energy
is examined.
In Section~\ref{sec:operators},
we give the low-energy behavior of
bulk and brane operators by the distance-rescaling.
We conclude in Section~\ref{conclusion} with
some remarks.
A five-dimensional Yukawa theory 
on the orbifold $S^1/Z_2$ is given 
in Appendix~\ref{sec:yukawa}.
The method we employ for
calculation of quantum loop corrections 
is exemplified 
in Appendix~\ref{sec:method}.
At one-loop level, bulk and brane divergences in
the five-dimensional Yukawa theory are found.
Details of mode functions and
their orthogonality and normalization are shown
in Appendix~\ref{app:mfg}.

\section{Model: brane kinetic terms and mode functions
\label{sec:model}}

We consider a theoretical improvement
of a five-dimensional Yukawa theory on the orbifold
$S^1/Z_2$.
The notation is given in Appendix~\ref{sec:yukawa}.
The theory has divergence on the branes.
The corresponding counterterms are needed.
We focus on the effect of brane terms
on the kinetic energy term.
Brane kinetic terms need to be included in the beginning,
\bea
  \int dy \, \left[ A \bar{\psi} (i\partial\!\!\!/)
 \psi + B \bar{\psi}_L
  (i\partial\!\!\!/) \psi_L
  \left(\delta(y)+\delta(y-L)\right) \right] ,
   \label{lagab}
\eea  
where the integral 
${1\over 2}\lim_{\epsilon\to 0} \int_{-L+\epsilon}^{L+\epsilon} dy$
is denoted as $\int dy$.
The factors $A$ and $B$ are unknown constants.
Either of these constants, for example, $A$ 
can be deleted by redefinition of $\psi$.
Then the equation (\ref{lagab}) reduces to
\bea
   \int dy\, \left[
   \bar{\psi} (i\partial\!\!\!/)
     \psi
   + a_0\, \bar{\psi}_L
     (i\partial\!\!\!/)\psi_L
       \left(\delta(y)+\delta(y-L)\right) \right].
       \label{lagka}
\eea
The coefficient $a_0$ should become momentum-dependent
after a renormalization.
The directly-related radiative effect is
fermion self-energy.
This
is given by the sum of bulk and brane contributions,
\bea
   \int {\cal D}\bar{\psi} {\cal D}\psi
     {\cal D}\phi \, \psi\bar{\psi} e^{i\int d^4 x 
       {\cal L}} ,
\eea
where the Lagrangian ${\cal L}$
consists of bulk and brane terms
and it is denoted 
as the four-dimensional effective Lagrangian
with a Kaluza-Klein decomposition.       
The basic idea to determine the coefficient $a_0$
is to require
that the propagator of the fermion 
with a physical mass
$M$ at a renormalization point is given  by
\bea
   {i\over p\!\!\!/ -M +i\epsilon}
   ={i\over p\!\!\!/ -M +i\epsilon} P_L
  + {i\over p\!\!\!/ -M +i\epsilon} P_R ,
   \label{idea}
\eea
where $P_L$ and $P_R$ denote projection 
matrices for left- and right-chiralities, respectively.
In the equation~(\ref{idea}), left and right contributions are required to be 
equal at the renormalization point.
As for the Lagrangian,
left and right components have different
terms as in Eq.~(\ref{lagka}) and 
the resulting radiative corrections
are expected to be different between left-
and right-chiralities.
It is the relative coefficient $a_0$ that
removes this difference by the renormalization,
while the common part of divergence 
is removed by the bulk renormalization constant.
We will show this occurs in the following sections.

To treat 
the renormalization,
we define 
the wave function 
renormalization factor and the 
rescaled field,
\bea
   \psi =Z^{1/2} \psi_r .
\eea
Substituting the rescaled field 
into the Lagrangian (\ref{lagka})
yields
\bea
 && \int
    dy \, \left[ Z \bar{\psi}_r (i\partial\!\!\!/) \psi_r
    + a_0 Z \bar{\psi}_{rL}
      (i\partial\!\!\!/) \psi_{rL}
        \left(\delta(y)+\delta(y-L)\right) \right]
\nonumber
\\
  &\!\!\!=\!\!\!&
  \int dy \, \left[
    \bar{\psi}_r (i\partial\!\!\!/)
   \psi_r
     +a\, \bar{\psi}_{rL}
      (i\partial\!\!\!/)
       \psi_{rL}
        \left(\delta(y)+\delta(y-L)\right) 
        \right.
\nonumber
\\
    && \left.
    + \delta_Z \bar{\psi}_r
      (i\partial\!\!\!/) \psi_r
      + \delta_a \bar{\psi}_{rL}
      (i\partial\!\!\!/)\psi_{rL}
      \left(\delta(y)+\delta(y-L)\right) 
    \right] ,
      \label{renaction}
\eea
where $a$ is a renormalized brane coupling
and the renormalization constants are
denoted as
$\delta_Z \equiv Z-1$ and
$\delta_a \equiv a_0 Z-a$.    
Hereafter we will omit the subscript $r$ to express 
the rescaled field.

We consider behavior of 
the coupling $a$ and the self-energy by loop effects for
the fermion in the Lagrangian
\bea
 {\cal L}&\!\!\!=\!\!\!& \int dy\, \bigg[
   \bar{\psi}(i\gamma^\mu \partial_\mu
     +i\gamma^5 \partial_5)\psi
     +{1\over 2}\left(\partial_\mu \phi\right)^2
     -{1\over 2}\left(\partial_5\phi\right)^2
     -g\bar{\psi}\psi\phi
\nonumber
\\
  && \qquad
  + a \, 
 \bar{\psi}_L (i\partial\!\!\!/) \psi_L
 \left(\delta(y)+\delta(y-L)\right)
  \bigg] ,
\eea
and the counterterms.
The equations of motion for the fermions are
\bea
   &&
      \partial_5 \psi_L + i\sigma^\mu \partial_\mu
        \psi_R = 0 ,
        \label{eom1}
\\
  &&
  i\bar{\sigma}^\mu
    \partial_\mu \psi_L
      -\partial_5 \psi_R
        +a \, i\bar{\sigma}^\mu \partial_\mu \psi_L
        \cdot \left(\delta(y)+\delta(y-L) \right) = 0.
   \label{eom2}
\eea
The mode expansion is given by 
\bea
   \psi_L (x,y) =
    \sum_{n=0}^\infty f_n (y) \psi_{Ln}(x) ,
 \qquad
  \psi_R (x,y) =
  \sum_{n=1}^\infty   g_n (y) \psi_{Rn}(x) .
  \label{modee}
\eea
The Dirac equations
$i\sigma^\mu \partial_\mu \psi_{Rn} (x) =
m_n \psi_{Ln}(x)$
and
$i\bar{\sigma}^\mu \partial_\mu \psi_{Ln}(x)
= m_n \psi_{Rn}(x)$ are fulfilled by
the four-dimensional fields.
The two equations have the identical $m_n$ 
for $n\geq 1$. 
Details of derivation of the
mode functions are given
in Appendix~\ref{app:mfg}.
The orthogonality is given by
\bea
 && \int_{0}^{L} dy \, f_n (y) f_m (y)
  + {a\over 2} \left( f_n(0) f_m(0)  
  +  f_n (L) f_m (L) \right)
    = \delta_{nm} ,
    \label{ortf}
\\ 
  && \int_0^L dy\, g_n (y) g_m(y)
     = \delta_{nm} .
      \label{ortg}
\eea
From these equations,
the Lagrangian for the fermion
is written in terms of 
the four-dimensional fields as
\bea
  && {1\over 2} \int_{-L +\epsilon}^{L+\epsilon}
   dy \, \left[ \bar{\psi}
     (i\gamma^\mu \partial_\mu + i\gamma^5 \partial_5)\psi
   +a \bar{\psi}_L (i\partial\!\!\!/) \psi_L
     (\delta(y) +\delta(y-L)) \right] 
\nonumber
\\
  &\!\!\!=\!\!\!&
   \chi_{L}^\dag i\bar{\sigma}\cdot \partial \chi_{L}
  +\sum_{n=1}^\infty 
  \left(
  \psi_{Ln}^\dag i\bar{\sigma}\cdot
   \partial \psi_{Ln}
   +\psi_{Rn}^\dag i\sigma\cdot \partial \psi_{Rn}
    -m_n \psi_{Rn}^\dag \psi_{Ln}
     -m_n \psi_{Ln}^\dag \psi_{Rn}
      \right) .
      \label{apluskk}
\eea
The kinetic terms are diagonal with respect to Kaluza-Klein modes
due to the orthogonalities of $f_n$ and $g_n$.

For the simplest five-dimensional Yukawa theory 
divergence for non-diagonal components 
with respect to
Kaluza-Klein modes are
radiatively generated, 
starting from diagonal kinetic terms.
Explicit equations are given in
Appendix~\ref{sec:method}.
The equation (\ref{apluskk}) is diagonal for
modes.
Also in the case with
brane kinetic terms,
radiative corrections are expected to 
give rise to
non-diagonal components.
To subtract this divergence,
the Lagrangian terms in 
the mode expansion for counterterms need to have non-diagonal
components with respect to Kaluza-Klein modes.
By the mode expansion,
the equation~(\ref{renaction}) is
\bea
  &&  \sum_n (\psi_{Rn}^\dag i\sigma \cdot \partial
      \psi_{Rn}
        + \psi_{Ln}^\dag i \bar{\sigma}\cdot \partial 
        \psi_{Ln})
\nonumber
\\
  && + \sum_n \delta_Z (\psi_{Rn}^\dag i\sigma \cdot \partial
      \psi_{Rn}
        + \psi_{Ln}^\dag i \bar{\sigma}\cdot \partial 
        \psi_{Ln})        
\nonumber
\\
  && 
    + \sum_m \sum_n
         (\delta_a -a \delta_Z)
           \psi_{Lm}^\dag
            i\bar{\sigma}\cdot \partial \psi_{Ln}
     \left(f_m(0) f_n(0)
         + f_m(L) f_n(L)\right) 
   \label{kincount} ,
\eea
where $\psi_{L0}\equiv \chi$.
For the rescaled fields (the subscript $r$ has been
omitted),
the kinetic terms are diagonal with respect to
Kaluza-Klein modes. 
In the last line,
the counterterms have off-diagonal components.
They are nonzero as we will see below.

For one-loop calculation,
the sum over mass for internal lines must be performed.
This is difficult when the mass eigenvalue
is given 
in terms of 
the mass quantization condition
in the form of a function.
Focusing on identifying effects of $a$,
we treat the fermion mass and scalar mass at 
the first order of $a$ as
\bea
  m_n = {n\pi \over L}
     \left(1-{a\over L}\right) + {\cal O}(a^2) ,
     \qquad
   m_n^\phi ={n\pi \over L} ,
\eea
respectively.
The corresponding mode functions are given by
\bea
   f_{n 
    }(y)
    &\!\!\!=\!\!\!& 
   \sqrt{2\over L}
        \left\{
          \left(1-{a\over 2L}\right)
            \cos \left({n\pi y\over L}\right)
         + {an\pi\over L}
           \left({y\over L}-{1\over 2}\right)
            \sin\left({n\pi y\over L}\right) \right\} 
   +{\cal O} (a^2) ,
\\
  g_{n 
   }(y)
    &\!\!\!=\!\!\!& \sqrt{2\over L}
      \left\{
        \left(1-{a\over 2L}\right)
          \sin\left({n\pi y\over L}\right)
           -{an\pi \over L}
           \left({y\over L} -{1\over 2}\right)
           \cos \left({n\pi y\over L}\right) \right\} 
           +{\cal O}(a^2) ,
\\
   g_n^\phi (y) &\!\!\!=\!\!\!& 
   \sqrt{2\over L}\sin\left({n\pi y\over L}\right) ,
\eea
for $0\leq y\leq L$.
From these mode functions,
the brane counterterm in Eq.~(\ref{kincount}) 
is proportional to
\bea
    f_{n 
       }(0)
    f_{m 
      }(0)
   +f_{n 
      } (L)
    f_{m 
      } (L)
  ={4\over L}
     \left(1-{a\over L}\right)
   {1+(-1)^{n+m}\over 2} 
   +{\cal O}(a^2) .
\eea
The off-diagonal components are nonvanishing when
the sum (or the difference) of $n$ and $m$
is an even number. 
 
In the present model, the only interaction
is the Yukawa interaction. 
The Yukawa coupling for the interaction
$(\bar{\psi}_{Ln}\psi_{Rm}\phi_\ell
+\bar{\psi}_{Rn}\psi_{Lm}\phi_\ell)$
in the four-dimensional Lagrangian is  
\bea
   (-g) \int_0^L dy \, 
   f_{n 
   }(y)
    g_{m 
   }(y)
    g_\ell^\phi (y)
 &\!\!\!=\!\!\!&
 {(-g)\over  \sqrt{2L}}
   \left(1-{a\over L}\right)
     \left(\delta_{\ell,m+n}
       +\delta_{\ell+n, m}
       -\delta_{\ell+m, n}\right)
\nonumber
\\
   && -{(-g)\over \sqrt{2L}}
   {a(n-m)\over L}
     \left[ 
       {1+(-1)^{\ell+n-m}\over 2(\ell+n-m)}
       + 
       {1+(-1)^{\ell-n+m}\over 2(\ell-n+m)}\right]
\nonumber
\\
  && \qquad \hspace{16ex}
            \ell+n\neq m
     \qquad \ell +m \neq n
\nonumber
\\
 &&
\nonumber
\\
  &&
   + {(-g)\over \sqrt{2L}}
      {a(n+m)\over L}
      \left[
        {1+(-1)^{\ell+n+m}\over 2(\ell+n+m)}
        + 
        {1+(-1)^{\ell-n-m}\over 2(\ell-n-m)} \right] 
\nonumber
\\
   && \qquad\hspace{34ex}
     n+m\neq \ell 
\nonumber
\\
  && + {\cal O}(a^2) ,
    \label{yukawacm}
\eea
where $\ell+n\neq m$ indicates that 
the term
$\left[1+(-1)^{\ell+n-m}\right]/
\left[2(\ell+n-m)\right]$ 
does not exist
for  $\ell+n= m$. The other indications
$\ell +m \neq n$ and $n+m\neq \ell$  are similar.

\section{One loop divergence and subtraction 
\label{sec:subtract}}

For the model given in the previous section,
we calculate fermion self-energy. 
The method is given 
in Appendix~\ref{sec:method}.
We discuss
correspondences between the divergences and the counterterms 
in the Lagrangian~(\ref{kincount}).

\subsection{Divergent part}

\begin{figure}[htb]
\begin{center}
\includegraphics[width=11cm]{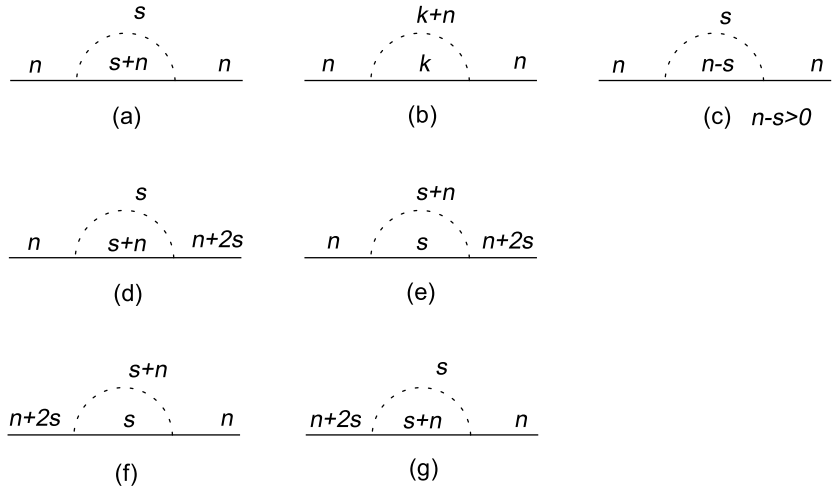}
\end{center}
\caption{Self-energy diagrams for $\psi_L$. 
\label{fig:kkl}}
\end{figure}

The first diagram we calculate is the one to have the
left-handed fermions with mode $n$
in the
external lines.
The diagrams are shown in Figure~\ref{fig:kkl}.
The internal lines are taken as $\psi_{Rm}$ and $\phi_\ell$.
The modes $m$ and $\ell$ are summed.
With the equation (\ref{yukawacm}),
the self-energy up to the propagators for the 
external lines is calculated
similarly to the one for
the diagrams in Figures~\ref{fig:chi} and \ref{fig:kkl}.
The divergence of $p\!\!\!/$ term
is found as    
\bea
    {g^2 \over 2L(4\pi)^2} \left(1-{a\over L}\right)
   (ip\!\!\!/)
   \left[{\Lambda L\over \sqrt{\pi}}
   -\log (\Lambda^2 L^2)  \right]
    \cdot P_L .
    \label{bulkdiva}
\eea
In the limit $a\to 0$, 
this reduces to the 
simplest-model
result (\ref{divabc}) for
$((\textrm{a})+(\textrm{b})+(\textrm{c}))$ in Figure~\ref{fig:kkl}.
The liner divergence $\Lambda L$ and
the logarithmic divergence $\log (\Lambda^2 L^2)$
correspond to bulk and brane terms, respectively. 
This is similar to Eq.~(\ref{simpletotal})
in the simplest model.
The divergence (\ref{bulkdiva}) is 
subtracted with a linear
combination
$\delta_Z$ and $\delta_a$
for the diagonal component in Eq.~(\ref{kincount}).
For the case with external right-handed fermions,
divergent part includes $a$ similarly to Eq.~(\ref{bulkdiva}).
For both of left- and right-handed fermions,
the bulk divergence needs to 
be subtracted with the identical counterterm.

The next diagram is
the one with $\psi_{Ln}$ and $\psi_{Ls}$ in the external lines
and with $\psi_{Rm}$ and $\phi_{\ell}$ in the internal lines
where the modes $m$ and $\ell$ are summed.          
In this case, the divergent part for $p\!\!\!/$ term 
is
\bea
    {g^2 \over 2L(4\pi)^2} (ip\!\!\!/)
      {1+(-1)^{n+s}\over 2}
       \left[
       -\left(1-{3a\over 2L}\right)
        \log(\Lambda^2 L^2)
        -{4a\over 3L}
        {\Lambda L\over \sqrt{\pi}} \right]\cdot P_L
        \label{aoff}
\eea
The linear divergence $\Lambda L$ appears 
as effects of non-zero $a$
because
the Yukawa coupling (\ref{yukawacm})
violates Kaluza-Klein number conservation at tree level.
To subtract this divergence,
the bulk kinetic terms in the original action integral seems to need
non-diagonal components with respect to Kaluza-Klein modes.
Introduction of the bulk off-diagonal components
might be accomplished when the five-dimensional
action integral is
regarded as an effective action in even higher dimensional theory.
While such a modification is possible,
it would be important to identify how the
bulk and brane kinetic terms in an effective theory 
vary their leading contributions
with the change of momentum scale.
At least, it is found that the divergences 
in the limit $a\to 0$ 
are completely subtracted with the counterterms
in Eq.~(\ref{kincount}).
We leave a further modification of the original action
for future work.

\subsection{Finite part}
So far we have focused on the divergent part in radiative corrections.
To obtain momentum dependence of wave function renormalizations,
the finite part need to be
derived. 

We consider the sum
$((\textrm{a})+ (\textrm{b}) + (\textrm{c}))$
in Figure~\ref{fig:kkl}.
From $\sum_{s=-\infty}^\infty$ term, 
we find
linear divergence in the bulk with a cutoff regularization.
This is the five-dimensional correspondent
of a four-dimensional logarithmic divergence.
On the other hand, 
the same four-dimensional divergence is expressed as
$\Gamma(2-d/2)|_{d= 4-\epsilon}=
\Gamma(\epsilon) \approx
 1/\epsilon$ in a dimensional regularization.
In five dimensions, $\Gamma(2-d/2)|_{d=5}=
\Gamma(-1/2)=
-2\sqrt{\pi}$ is finite.
The momentum integral seems regularization-dependent.
If physical quantities are sensitive to the way of the regularization, the
predictability would be lost.
In the present case, it will be found that
regularization-dependent divergent terms
are independent of the momentum scale up to the overall
$p\!\!\!/$.
Such a momentum-independent shift 
is unphysical for the renormalization.
In Eq.~(\ref{sumabc}), the sum part is written as
\bea
    &&
     {g^2 \over 2L}
       \sum_{s=-\infty}^\infty
       \int_0^1 dx
       \int {d^4\ell\over (2\pi)^4}
       \left(
       {1\over \left[\ell^2 -\Delta_{s+}\right]^2}
       -{1\over \left[\ell^2 -\Delta_{s+}|_{p^2=0}\right]^2}
       \right)
        x p\!\!\!/ P_L
\nonumber
\\
  &&  
   +{g^2 \over 2L}\sum_{s=-\infty}^\infty
     \int_0^1 dx \int {d^4 \ell\over (2\pi)^4}
       {xp\!\!\!/\over 
       \left[\ell^2 -\Delta_{s+}|_{p^2=0}\right]^2}
       P_L .
\eea
Only the last term diverge and it corresponds to a constant
shift.
As seen in Eq.~(\ref{simpletotal}),
all the divergence for this part and the others can be 
subtracted with only the two factors
$\delta_Z$ and $\delta_a$. 
In the contribution 
$((\textrm{a})+(\textrm{b})+
(\textrm{c}))$, the momentum-dependent part is
obtained as
\bea
 -{g^2 \over 2L(4\pi)^2}
     (ip\!\!\!/)
     \int_0^1 dx \, x
        \left[
          -\log(\Delta_{0+} L^2)
          -\log(\Delta_{n+} L^2)
     +\sum_{s=-\infty}^\infty
        \log \left({\Delta_{s+}
          \over \Delta_{s+}|_{p^2=0}}\right)
           \right] \, P_L .
\eea
For the external $\psi_n$ and $\psi_{n+2s}$,
the self-energy $((\textrm{d})+(\textrm{e}))$
is
\bea
   {g^2 \over 2L(4\pi)^2}
      (ip\!\!\!/)
        \int_0^1 dx \, \log
          (\Delta (m_{n+s}, m_s)L^2) \cdot P_L ,
\eea
where only the momentum-dependent part has been given.
The self-energy $((\textrm{f})+(\textrm{g}))$ 
has the same value.
For the right-handed external lines,
the momentum-dependent part is
\bea
  && -{g^2 \over 2L(4\pi)^2}
     (ip\!\!\!/)
     \int_0^1 dx \, x
       \sum_{s=-\infty}^\infty
        \log \left({\Delta_{s+}
          \over \Delta_{s+}|_{p^2=0}}\right)
           \cdot P_R .
\eea
In the next section, we will examine
momentum dependence of wave function renormalizations
for bulk and brane terms 
at the leading level.

\section{Momentum dependence of self-energy
\label{sec:momentum}}

Now we perform the renormalization.
The full propagator $G_{n,\ell}$
is related to the one-particle irreducible self-energy
$\Sigma_{n,\ell}$ and the tree propagator $S_{n,n}$
as 
\bea
   G_{n,n} = S_{n,n} + \sum_j
     S_{n,n} \Sigma_{n,j} G_{j,n} ,
     \qquad
   G_{n,\ell} =\sum_{k} S_{n,n} 
     \Sigma_{n,k} G_{k,\ell} , 
 \quad n\neq \ell
\eea
where $\Sigma_{n,j} = \Sigma_{j,n}$
and $G_{n,\ell} = G_{\ell,n}$.
The solution at the leading level is given by
\bea
  G_{n,n} ={S_{n,n} \over 1-S_{n,n} \Sigma_{n,n}}
  ,
  \qquad
  G_{n,\ell} = S_{n,n} \Sigma_{n,\ell} S_{\ell,\ell}
  .
\eea
The diagonal and off-diagonal parts are separated.
From the Lagrangian (\ref{kincount}),
the renormalized self-energy $-i\hat{\Sigma}$, 
the one-loop self-energy $-i\Sigma^{(1)}$ and the
counterterms are related to each other as follows:
\bea
   -i\hat{\Sigma}_{L\, n,n} (p\!\!\!/) P_L
   &\!\!\!=\!\!\!&
     -i\Sigma_{L\, n,n}^{(1)} (p\!\!\!/) P_L
       +ip\!\!\!/ \delta_Z P_L
       +ip\!\!\!/ (\delta_a -a\delta_Z){2\over L}  P_L,
   \label{hatL}
\\
   -i\hat{\Sigma}_{R\, n,n} (p\!\!\!/) P_R
     &\!\!\!=\!\!\!&
       -i\Sigma_{R\, n,n}^{(1)} (p\!\!\!/) P_R
         +ip\!\!\!/ \delta_Z  P_R ,
         \label{hatR}
\\
   -i\hat{\Sigma}_{L \, n+2s,n} (p\!\!\!/) P_L
     &\!\!\!=\!\!\!&
       -i\Sigma_{L\, n+2s, n}^{(1)} (p\!\!\!/) P_L
         +ip\!\!\!/ (\delta_a -a \delta_Z) 
           {2\over L} P_L,
\\
   -i\hat{\Sigma}_{L\, n,n+2s}(p\!\!\!/) P_L
     &\!\!\!=\!\!\!&
   -i\Sigma_{L \, n,n+2s}^{(1)}(p\!\!\!/) P_L
 +ip\!\!\!/  (\delta_a -a \delta_Z){2\over L} P_L,
\eea         
where the subscripts 
$L$ and $R$ label left- and right-handed
chiralities for external fermion $\psi$
with leaving a specification for anti-fermion 
$\bar{\psi}$, respectively.
Physical conditions for the self-energies
are given in the following.
For the propagator for 
fermion with
a physical mass
to keep the factor in the numerator $i$,
the renormalization conditions for the diagonal part are
imposed as
\bea
  \left.
  {d\hat{\Sigma}_{L \, n,n}\over d p\!\!\!/}\right|_{p\!\!\!/ =m_n}
    =0 ,
  \qquad
    \left.
  {d\hat{\Sigma}_{R \, n,n}\over d p\!\!\!/}\right|_{p\!\!\!/ =m_n}
    =0 ,
    \label{conditions}
\eea
where $m_n$ is the physical mass for the propagator
\bea
  i \left[
  p\!\!\!/-m_n -\hat{\Sigma}_{L \, n,n}(p\!\!\!/)
   \right]^{-1} P_L
  =
   i \left[ 
  (p\!\!\!/-m_n)(1-\hat{\Sigma}'_{L\, n,n}(m_n))
+{\cal O}((p\!\!\!/-m_n)^2) \right]^{-1} P_L,
  \label{prop}
\eea
with the same form of equation for 
$\hat{\Sigma}_{R\, n,n}$.
From the condition for $\hat{\Sigma}_{R\, n,n}$ in
Eq.~(\ref{conditions}), 
the counterterm $\delta_Z$ is obtained as
\bea
  \delta_Z = \left.
   {d \Sigma_{R\, n,n}^{(1)} \over dp\!\!\!/}  
      \right|_{p\!\!\!/=m_n= p_5} , 
      \label{deltaz}
\eea
where $p_5$ is the Fourier transformation of extra-dimensional 
derivative.
The equation for the counterterm should be interpreted
in terms of $p_5$ rather than the form dependent on Kaluza-Klein
modes.
From this equation and the condition for
the left-handed component in Eq.~(\ref{conditions}),
the other counterterm is obtained as
\bea
    (\delta_a -a \delta_Z){2\over L}
       = \left.
        {d\over dp\!\!\!/}
           \left(\Sigma_{L\, n,n}^{(1)}
             (p\!\!\!/)
         -\Sigma_{R\, n,n}^{(1)}(p\!\!\!/) \right) 
         \right|_{p\!\!\!/=m_n=p_5}.
         \label{deltaa}
\eea
The renormalization conditions for the off-diagonal external 
lines
$n+2s,n$ and $n, n+2s$ need to be imposed as
$\hat{\Sigma}_{L\, n+2s,n}(p_5)
 =\hat{\Sigma}_{L\, n,n+2s}(p_5)
 = 0$,
so that the field for the propagator
(\ref{prop}) is in the mass eigenstate. 
In these equations, the Kaluza-Klein mass-dependent
part
in $\Sigma_{L\, n+2s, n}^{(1)}$ and 
$\Sigma_{L\, n+2s, n}^{(1)}$
should be read with the Fourier-transformed derivative $p_5$.

These self-energies can be decomposed into bulk and 
brane parts.
The bulk part is defined from the contribution for
the right-handed fermion as
\bea
  {g^2\over 2L(4\pi)^2}
        p\!\!\!/ \,\Sigma_{\textrm{\scriptsize bulk}}^{(1)}(p\!\!\!/) =
   \Sigma_{R\,n,n}^{(1)}
   (p\!\!\!/)       .
          \label{sigbulk}
\eea
Here the correspondent with hat
is defined similarly.
From the equations (\ref{hatR}), (\ref{deltaz})
and (\ref{sigbulk}), 
the renormalized $\hat{\Sigma}_{\textrm{%
\scriptsize bulk}}$ is
\bea
   \hat{\Sigma}_{\textrm{\scriptsize bulk}} (p\!\!\!/)
  =
    \int_0^1 dx \, x \sum_{s=-\infty}^\infty
   \left[  
   \log\left(
       1 + {x(1-x)(p_5^2 -p^2)L^2
        \over (s\pi + (1-x)p_5 L)^2}\right)
    +
      {2x (1-x) p\!\!\!/ p_5 L^2 \over
      (s\pi +(1-x)p_5 L)^2} 
    \right] .
      \label{sbue}
\eea
Next the brane part is defined 
from the difference between the contributions
for the left-handed and right-handed fermions as
\bea
  {g^2 \over 2L(4\pi)^2}
   p\!\!\!/ \, \Sigma_{\textrm{\scriptsize brane}}^{(1)} (p\!\!\!/)
  = \Sigma_{L\, n,n}^{(1)} (p\!\!\!/)
   -\Sigma_{R\, n,n}^{(1)} (p\!\!\!/) .
   \label{sigbrane}
\eea   
Here the correspondent with hat
is defined similarly. 
From the equations (\ref{hatL}), (\ref{hatR}),
(\ref{deltaa}) and (\ref{sigbrane}),
the renormalized $\hat{\Sigma}_{\textrm{\scriptsize
brane}}$ is
\bea
  \hat{\Sigma}_{\textrm{\scriptsize brane}}
    (p\!\!\!/)
  &\!\!\!=\!\!\!&
   \int_0^1 dx \left( \log
   \left({-x(1-x)p^2 +(1-x) p_5^2 \over
   (1-x)^2 p_5^2} \right)
  \right.
\nonumber
\\
  && \left. 
  +\log \left(
    { -x(1-x) p^2 +(4-3x)p_5^2 \over
    (2-x)^2 p_5^2 }
     \right)
      \right)
\nonumber
\\
  &&
    +  \int_0^1 dx
         \, 2x^2 (1-x) p\!\!\!/ p_5
         \left(
           {1\over (1-x)^2 p_5^2}
             +{1\over (2-x)^2 p_5^2}
     \right)  .
       \label{sbre}
\eea
Thus the self-energies (\ref{sbue}) and
(\ref{sbre}) have been determined uniquely.

The bulk and brane parts of the self-energy
is numerically estimated.
For the equation
(\ref{sbue}) and (\ref{sbre}),
the self-energies
are singular for $x=1$.
This is the same as in infrared divergence 
for massless photon in the 
four-dimensional quantum electrodynamics.
If the five-dimensional scalar field is massive,
this infrared divergence is avoided.
For the numerical analysis,
we take the values of 
$p_5$ and the five-dimensional scalar
mass as the typical dimensional quantity $1/L$.
By defining
$Q \equiv (p\!\!\!/L)/(p_5 L)$,
we examine the momentum dependence of
the bulk and brane self-energies.
\begin{figure}[htb]
\begin{center}
\includegraphics[width=10cm]{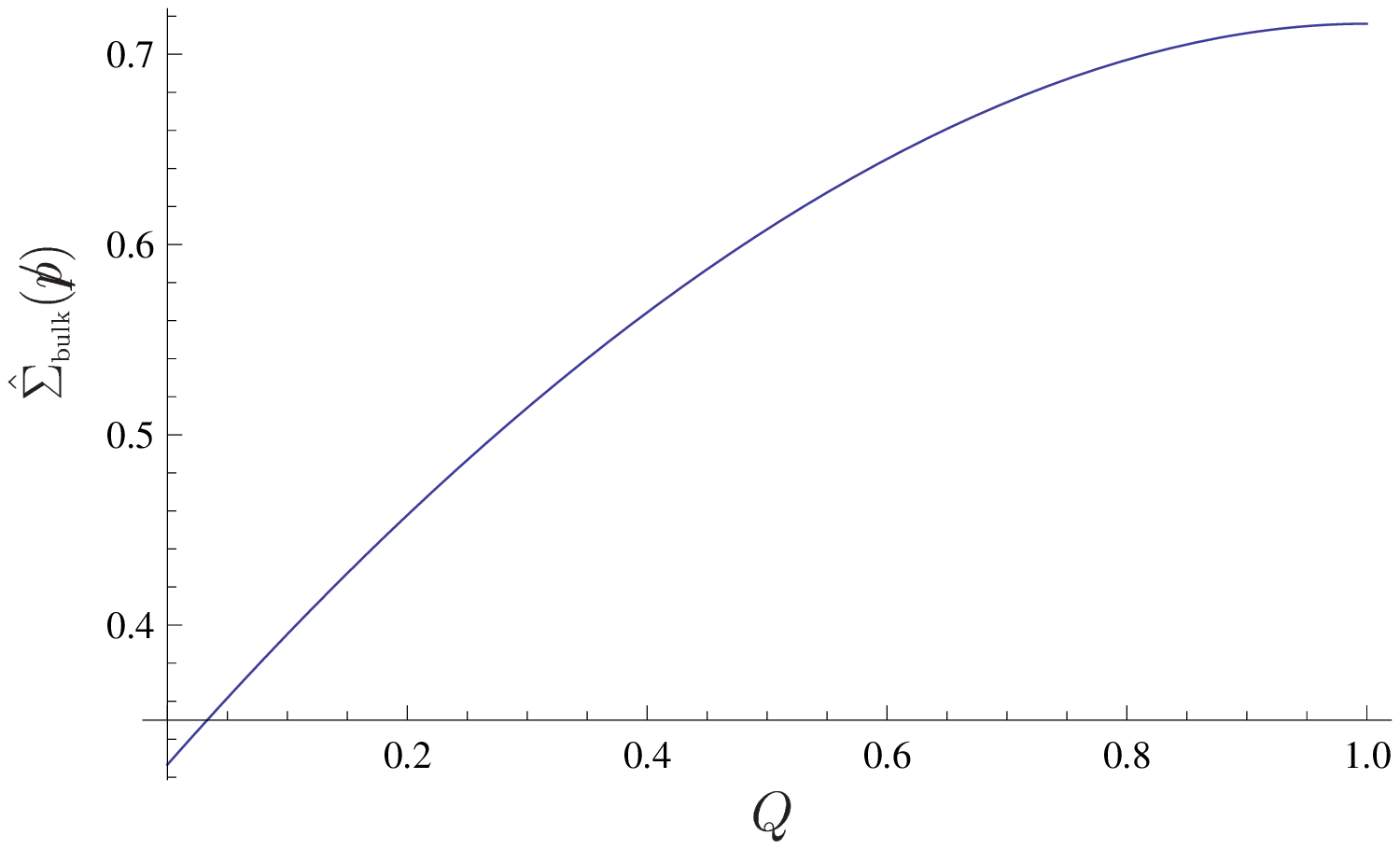}
\end{center}
\caption{The momentum dependence of $\hat{\Sigma}_{\textrm{\scriptsize bulk}}(p\!\!\!/)$,
where $p_5=1/L$ and $-10^3 \leq s \leq 10^3$.
\label{fig:sigma}}
\end{figure}
The momentum dependence of
$\hat{\Sigma}_{\textrm{\scriptsize bulk}}(p\!\!\!/)$
is shown
in Figure~\ref{fig:sigma}.
For modes of the Poisson summation,
we take the sum $-10^3 \leq s \leq 10^3$
as
the contribution for $|s| > 10^3$ is much less than
${\cal O}(\%)$.
For $Q\sim 0.4$, the linear and logarithmic 
parts of Eq.~(\ref{sbue}) are comparable.
For $0.7\siml Q \leq 1$, the linear part
is dominant.
For $Q=1$, i.e., $p\!\!\!/ =p_5$,
the logarithmic part is vanishing.
The renormalized bulk self-energy is written as
the linear term and the others such as
$\hat{\Sigma}_{\textrm{\scriptsize bulk}}
\approx 0.7 Q + \hat{\Sigma}_{%
\textrm{\scriptsize bulk}}^{%
\textrm{\scriptsize nonlinear}}$,
where $\hat{\Sigma}_{\textrm{\scriptsize
bulk}}|_{Q=1}\simeq 0.7$.
For $10^{-5} \leq Q \leq 1$,
the bulk self-energy
$\hat{\Sigma}_{\textrm{\scriptsize bulk}}$
changes the value by about 50$\%$.
\begin{figure}[htb]
\begin{center}
\includegraphics[width=10cm]{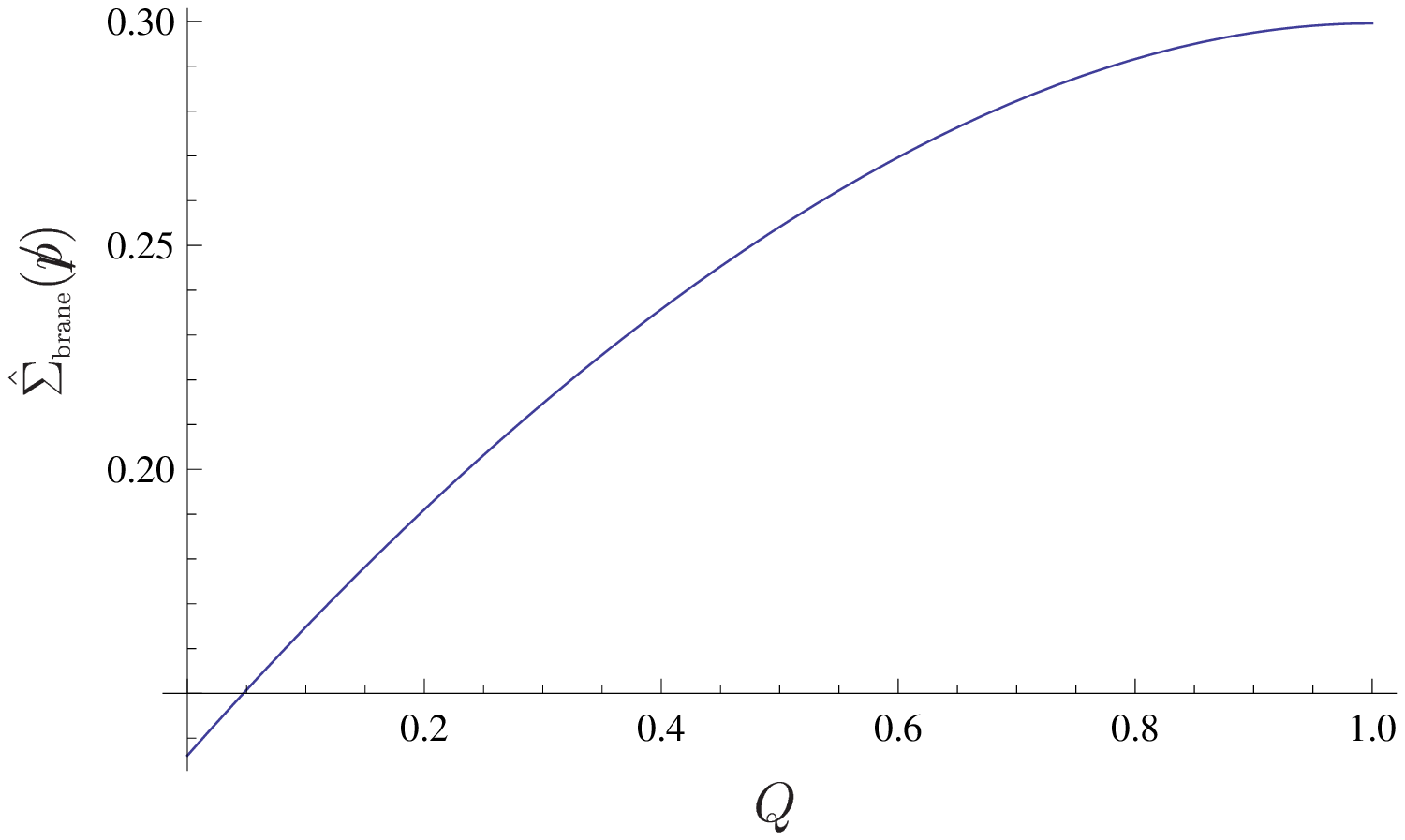}
\end{center}
\caption{The momentum dependence of
$\hat{\Sigma}_{\textrm{\scriptsize
brane}}(p\!\!\!/)$,
where the value for $p_5$ is
taken as the same as in Figure~\ref{fig:sigma}.
\label{fig:sigma2}}
\end{figure}
The momentum dependence of 
$\hat{\Sigma}_{\textrm{\scriptsize brane}}(p\!\!\!/)$
is shown in Figure~\ref{fig:sigma2}.
The behavior is similar to the case of 
$\hat{\Sigma}_{\textrm{\scriptsize bulk}}$.
The brane self-energy is written as
the linear term and the others such as
$\hat{\Sigma}_{\textrm{\scriptsize brane}}
\approx 0.3 Q + \hat{\Sigma}_{%
\textrm{\scriptsize brane}}^{%
\textrm{\scriptsize nonlinear}}$,
where $\hat{\Sigma}_{\textrm{\scriptsize
brane}}|_{Q=1}\simeq 0.3$.
In addition to each momentum-dependence of
the bulk and brane parts, the momentum-dependence
of the ratio is important.
The ratio of the brane contribution to the              
bulk contribution is shown in Figure~\ref{fig:sratio}.
\begin{figure}[htb]
\begin{center}
\includegraphics[width=10cm]{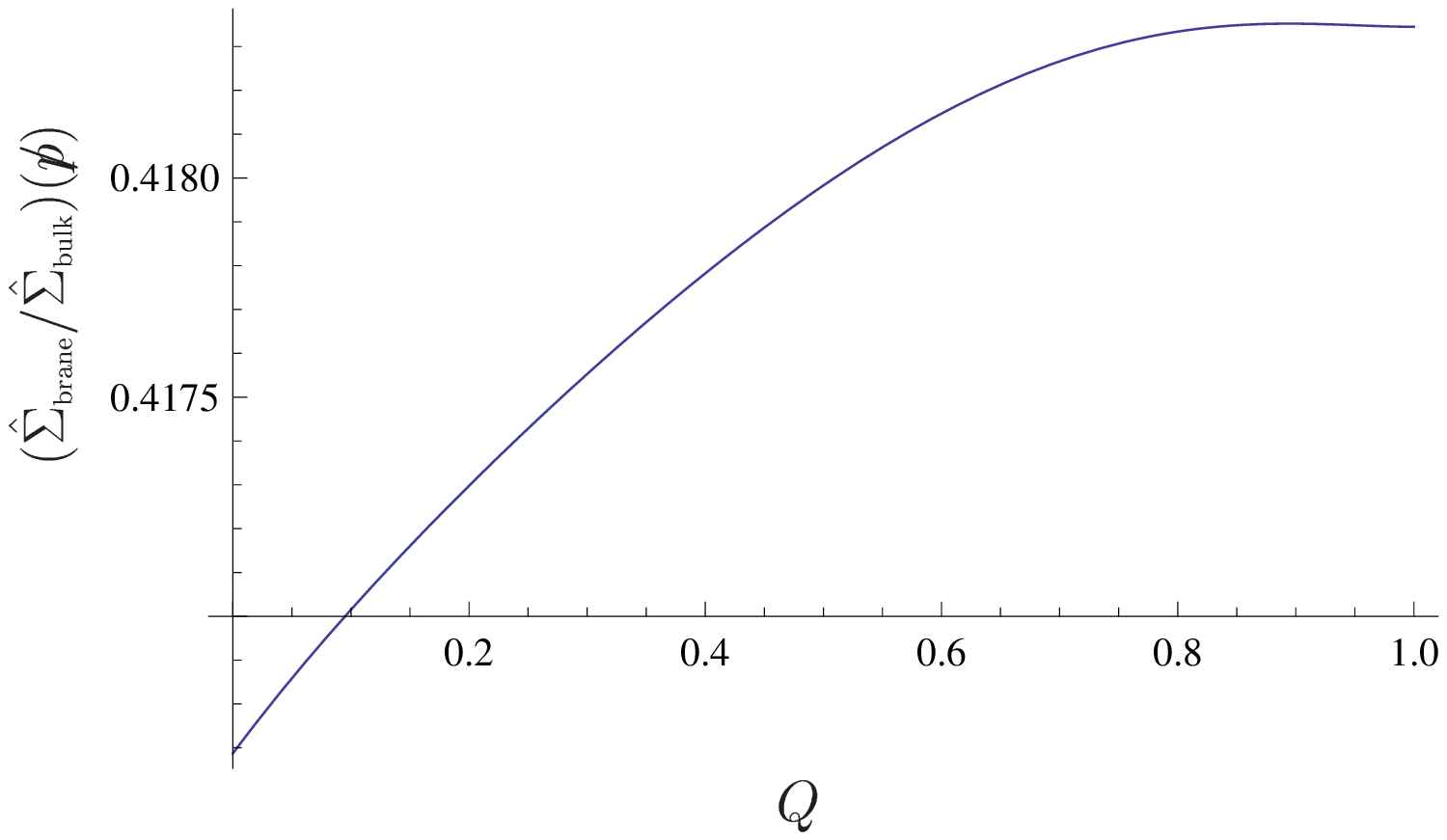}
\end{center}
\caption{The ratio
$(\hat{\Sigma}_{\textrm{\scriptsize brane}}
/\hat{\Sigma}_{\textrm{\scriptsize
bulk}})(p\!\!\!/)$, where the value for 
$p_5$ is 
taken as
the same as in Figure~\ref{fig:sigma}.
\label{fig:sratio}}
\end{figure}
For $0.7 \siml Q\leq 1$,
the ratio is dominated by the linear term
and is approximated as
$\hat{\Sigma}_{\textrm{\scriptsize brane}}
/\hat{\Sigma}_{\textrm{\scriptsize bulk}}
\approx 0.4 Q + (\textrm{nonlinear})$,
where $0.3/0.7\simeq 0.4$.
The nonlinear terms have logarithmic dependence
of $Q$ and the change of the value against the 
momentum-scale is small.
For $10^{-5} \leq Q \leq 1$,
each of $\hat{\Sigma}_{\textrm{\scriptsize bulk}}$
and $\hat{\Sigma}_{\textrm{\scriptsize brane}}$
changes the value by about 50$\%$.
On the other hand,
the change of the value for the ratio is 
less than 0.1$\%$.
The momentum insensitivity of the ratio
is due to
 the analogous form seen in $\hat{\Sigma}_{\textrm{\scriptsize bulk}}$
and $\hat{\Sigma}_{\textrm{\scriptsize brane}}$.

From the result for the self-energy, 
it follows that
the ratio of the brane contribution to
the bulk contribution is almost constant
with respect to the momentum-scale.
When $(\delta_a -a\delta_Z) /\delta_Z$
is constant, the $\delta_a$ is represented in terms
of $\delta_Z$ as
$\delta_a = (a + \textrm{constant}) \delta_Z$.
At the leading correction, $a=0$.
Then the Lagrangian (\ref{renaction})
can be represented as
\bea
  \int dy \, \left[
    \bar{\psi}_r (i\partial\!\!\!/)
   \psi_r
    + \delta_Z
   \left\{ \bar{\psi}_r
      (i\partial\!\!\!/) \psi_r
      + 0.42 \, \bar{\psi}_{rL}
      (i\partial\!\!\!/)\psi_{rL}
      \left(\delta(y)+\delta(y-L)\right) 
      \right\}
    \right] ,
\eea
where 0.42 is the value of the ratio
for the region
in Figure~\ref{fig:sratio}.
Therefore the relative coefficient between
the bulk and brane kinetic terms
has been determined without ambiguity.

\section{Bulk and brane operators in
the distance-rescaling 
\label{sec:operators}}
In this section, we discuss the issue of 
relevant, irrelevant and marginal operators for
bulk and brane terms to identify a general low-energy
behavior of quantum corrections
in the field-theoretical context.

In Ref.~\cite{ue},
marginal and interacting operators 
in quantum field theory with extra dimensions
were discovered for the Randall-Sundrum
spacetime whose metric is given by~\cite{%
Randall:1999ee, 
Randall:1999vf}
\bea
  ds^2 = {1\over z^2}
    \left(\eta_{\mu\nu} dx^\mu dx^\nu
      -{1\over k^2} dz^2\right) ,
\eea
where $k$ is the curvature of the five-dimensional
anti-de Sitter space.
For the four-dimensional rescaling 
$x'=bx$ where $b<1$, the rescaling of
$z$ is given by
$z'=z/b$.
The action integral
\bea
  \int d^4 x {dz \over kz}
    \left[ {1\over z^3}
   \bar{\psi} i\gamma^\mu \partial_\mu \psi 
  + {1\over 2 z^2} (\partial_\mu \phi)^2
      -{1\over  z^4} g\bar{\psi}\psi \phi
   \right] ,   
\eea
is rescaled into 
\bea
  \int d^4 x' {dz' \over kz'}
    \left[ {1\over {z'}^3}
   \bar{\psi}' i\gamma^\mu \partial'_\mu \psi' 
  + {1\over 2 {z'}^2} (\partial'_\mu \phi')^2
      -{1\over {z'}^4} g' \bar{\psi}' \psi' \phi'
   \right] .   
\eea
Here the field redefinition is
given by $\psi'=b^{-3}\psi$
and $\phi'= b^{-2} \phi$.
The coupling constant is given
by $g'= b^0 g$ 
and the corresponding interaction is
a marginal operator.

If the spacetime is flat with
the rescaling 
$x'=bx$ and $y' =by$,
the action integral
\bea
  \int d^4 x dy
    \left[ 
   \bar{\psi} i\gamma^\mu \partial_\mu \psi 
  + {1\over 2} (\partial_\mu \phi)^2
      -g \bar{\psi}\psi \phi
   \right] ,   
\eea
is rescaled into
\bea
  \int d^4 x dy
    \left[ 
   \bar{\psi}' i\gamma^\mu \partial'_\mu \psi' 
  + {1\over 2} (\partial'_\mu \phi')^2
      -g \bar{\psi}' \psi' \phi'
   \right] .
\eea
Here the field redefinition is
given by $\psi'=b^{-2} \psi$ 
and $\phi' = b^{-3/2}\phi$.
The coupling constant is given by
$g'= b^{1/2} g$ and
the corresponding operator is an irrelevant operator.
 
Now we consider the sum of bulk and brane terms.
In the flat spacetime, the action integral is 
given by
\bea
  \int d^4x dy \,
   {\cal L}_{\textrm{\scriptsize bulk}}
    (\psi,\phi)
   + \int d^4 x dy \,
   {\cal L}_{\textrm{\scriptsize brane}}
   (\psi,\phi) 
 \delta(y-y_i),
   \label{actionflat}
\eea
where $y_i$ denote the positions of branes. 
For the rescaling in the flat spacetime,
the action integral is written as
\bea
  \int d^4x' dy' b^{-5} \,
   {\cal L}_{\textrm{\scriptsize bulk}}
    (b^2 \psi',b^{3/2} \phi')
   + \int d^4 x' dy' b^{-5} \,
   {\cal L}_{\textrm{\scriptsize brane}}
   (b^2 \psi',b^{3/2} \phi') 
  \delta(y'-y'_i) b.
\eea
The brane terms
are multiplied by the factor $b$ compared to
the bulk terms.
This shows that in the flat spacetime
brane kinetic terms are
small compared to bulk kinetic terms.

In the Randall-Sundrum spacetime,
the action integral corresponding to 
Eq.~(\ref{actionflat}) is given by
\bea
  \int d^4x dz \sqrt{\textrm{det}g_{MN}}
   {\cal L}_{\textrm{\scriptsize bulk}} (\psi, \phi ;
     g_{MN})
   + \int d^4 x dz \sqrt{\textrm{det}g_{MN}}
   {\cal L}_{\textrm{\scriptsize brane}} 
  (\psi, \phi ; g_{MN}) kz \delta(z-z_i).
\eea
For the rescaling in the Randall-Sundrum spacetime,
the action integral is written as
\bea
 && \int d^4x' dz' \sqrt{\textrm{det}g'_{MN}}
   b^{-8}
   {\cal L}_{\textrm{\scriptsize bulk}} 
  (b^{3}\psi', b^{2}\phi' ;
     b^{-2} g'_{MN})
\nonumber
\\
 &&  + \int d^4 x' dz' \sqrt{\textrm{det}g'_{MN}}
   b^{-8}
   {\cal L}_{\textrm{\scriptsize brane}} 
  (b^{3} \psi',b^{2} \phi' ; b^{-2} g'_{MN}) 
  kz' \delta(z'-z'_i) .
\eea
Because $kz \delta(z-z_i) =kz' \delta(z'-z'_i) b^0$,
the brane kinetic terms are the same order as 
the bulk kinetic terms.

Our method to determine the momentum-dependence
of the bulk and brane contributions has been given
in the flat spacetime.
In the flat spacetime, the bulk interactions
and the brane kinetic terms tend to be irrelevant
operators. 
Thus our explicit analysis may
be regarded as the proposal of an idea in a simplified
model.
On the other hand,
in the Randall-Sundrum spacetime,
these operators tend to be marginal operators.
In the four-dimensional theory, marginal operators
are renormalizable terms and low-energy
quantum corrections
are consistently treated without an additional
ultraviolet completion.
The idea needs to be examined further in 
models where the bulk interactions and 
the brane kinetic terms are marginal operators.

\section{Conclusion \label{conclusion}}

We have examined
energy-scale dependence of fermion self-energy
associated with wave
function renormalization
in Yukawa theory
in flat five dimensions on the $S^1/Z_2$.
At one-loop level, wave function renormalization 
is needed as in four dimensions.
One-loop divergence can be subtracted
with only the two renormalization constants
$\delta_Z$ and $\delta_a$
in the bulk and on the branes.
We have shown that the bulk and brane parts of the
self-energies are uniquely determined with
requiring physical conditions.
The physical condition is that
fields with their physical mass obey 
usual Feynman propagators.
In the present setup,
for the right-handed component
the bulk part of the self-energy is immediately
fixed.
The left-handed component determines the other
renormalization factor.
With this procedure,
momentum-scale dependence of the 
renormalized self-energies
has been given.
While the loop integral has 
linear divergence in the cutoff regularization,
the renormalized self-energy
does not give rise to 
regularization-scheme dependence.
As an explicit equation,
the bulk part of the renormalized 
self-energy is rewritten as
\bea
  \hat{\Sigma}_{\textrm{\scriptsize
  bulk}}
   (p\!\!\!/)
    =\Sigma_{\textrm{\scriptsize bulk}}^{(1)}
       (p\!\!\!/)
      - \Sigma_{\textrm{\scriptsize bulk}}^{(1)}
         (p_5)
     -p\!\!\!/ \left.
        \left({d\over dp\!\!\!/}
         \Sigma_{\textrm{\scriptsize 
         bulk}}^{(1)}(p\!\!\!/)
         \right)\right|_{
          p\!\!\!/=p_5} .       
\eea
From this equation,   
it is seen that 
a constant shift of $\hat{\Sigma}_{\textrm{
 \scriptsize bulk}}^{(1)}$ is unphysical.
The case of $\hat{\Sigma}_{\textrm{\scriptsize
  brane}}$ is similar.
We have also found  that
the ratio of the bulk
contribution to the brane contribution
depends on the momentum scale within
less than 0.1$\%$,
so that
the relative coefficient of 
bulk and brane kinetic terms
can be regarded as 
approximately constant for the leading
quantum effect.
The value of the relative coefficient
has been estimated as 0.42 
for the parameter set given 
in Section~\ref{sec:momentum}.
Because the value is not extremely
small, it would affect the prediction
in phenomenological applications such as
collider physics.
The quantitative estimation depends on 
the model.

Analysis
beyond the leading level is an open question.
In the equation (\ref{aoff}), divergence for
the off-diagonal component with respect to Kaluza-Klein
modes has been found.
To subtract this divergence,
the original action integral would need
the corresponding counterterms. 
As mentioned below Eq.~(\ref{aoff}),
linear divergence for the off-diagonal component 
may correspond to five-dimensional terms
in higher-dimensional spacetime.
The necessity of new local operators order by order
is nothing but non-renormalizability.
A possible way to proceed is to show
that such local operators are irrelevant.
In the present analysis,
the relative coefficient of the brane terms to 
the bulk term is 0.42 which is not negligible.
The effects of new operators should be treated
carefully.
As for regularization-scheme dependence,
it needs to be checked whether the 
scheme independence is
kept at 
higher-loop level. 
Because the loop-integral is scheme dependent
even at the leading level,
the momentum-dependent part 
for any physical quantities
might be contributed from
such a effect.
A viewpoint of this issue is given in the following.
When our interest is low energy behavior of 
physical quantities, 
details of the scheme-dependent part
may not be required.
Then we do not need to 
show that the prediction is exactly scheme independent.
Rather, it is sufficient that the
scheme-dependence is negligible 
at low energies.
In addition,
our analysis has focused on 
wave function renormalization
for two-point functions.
To define a physical mass, 
mass renormalization should be treated appropriately.

Finally we emphasize that
the physical conditions given here
are applicable to
remove ambiguity 
in various orbifold models.
We have also shown that
both of the bulk and brane kinetic terms are 
marginal operators in the Randall-Sundrum spacetime.
It would be worthwhile to 
identify the momentum-dependence of 
bulk and brane contributions
in such models.

\vspace{8ex}

\subsubsection*{Acknowledgments}

This work is supported by Scientific Grants 
from the Ministry of Education
and Science, Grant No.~20244028.

\newpage

\begin{appendix}

\section{Yukawa theory on an orbifold
\label{sec:yukawa}}

We consider
a five-dimensional 
Yukawa theory on the orbifold 
$S^1/Z_2$. The bulk 
spacetime is flat with the metric
$\eta^{MN}=\textrm{diag}(1,-1,-1,-1,-1)$.
The action integral for
the Dirac fermion $\psi(x,y)$ 
and the real scalar field $\phi(x,y)$ is
\bea
   S = \int d^4 x \, \cdot
  {1\over 2} \int_{-L}^L
    dy \, \left[\bar{\psi}
  (i\gamma^\mu \partial_\mu +i\gamma^5 \partial_5
  )\psi 
  +{1\over 2} (\partial_\mu \phi)^2
  -{1\over 2} (\partial_5 \phi)^2
   -g\bar{\psi}\psi\phi \right].
    \label{actiony}
\eea
Greek indices $\mu$ run over
0,1,2,3 and fifth index is denoted as $y$.
The gamma matrices are
\bea
   \gamma^\mu = \left(\begin{array}{cc}
      & \sigma^\mu \\
      \bar{\sigma}^\mu & \\
       \end{array}\right)
       , \qquad
   -i\gamma^5 =\left( \begin{array}{cc}
  -{\bf 1}_2 & \\
     & {\bf 1}_2 \\
   \end{array}\right) 
  =i\gamma_5 ,
\eea
where $\sigma^\mu =({\bf 1}_2,\sigma^i)$ and 
$\bar{\sigma}^\mu=(-{\bf 1}_2,\sigma^i)$ with
the Pauli matrices.
Dirac conjugate is given by
$\bar{\psi} =\psi^\dag \gamma^0$.
The fields have period $2L$ with respect to 
the extra dimension.
The orbifold boundary conditions for 
the fields are chosen as
\bea
    \psi (x, -y) &\!\!\!=\!\!\!&
   i\gamma^5 \psi (x,y) , 
  \qquad
   \psi (x, L-y) =i\gamma^5 \psi(x,L+y) , 
    \label{psiLR}
\\
     \phi(x,-y) &\!\!\!=\!\!\!&
   -\phi (x,y) ,\qquad
     \phi(x,L-y) =-\phi (x,L+y) ,
\eea 
for the extra-dimensional fundamental region
$0\leq y \leq L$.
With this boundary conditions,
the fermion has
mode expansion as
\bea
   \psi_L (x,y )&\!\!\!=\!\!\!&
     {1\over \sqrt{L}} \,\chi_{L} (x) 
        + \sum_{n=1}^\infty \sqrt{2\over L} \,\psi_{Ln} (x)
          \cos \left( 
     m_n y \right) ,
       \label{psil}
\\
   \psi_R(x,y) &\!\!\!=\!\!\!&
     \sum_{n=1}^\infty
      \sqrt{2\over L} \,\psi_{Rn}(x)\sin\left(
     m_n y \right) .
         \label{psir}
\eea
with $m_n =n\pi /L$.
The projection matrices
are given by 
$P_L =\left[{\bf 1}_2-(-i\gamma^5)\right]/2 =\textrm{diag}(1,1,0,0)$ 
and $P_R =\left[{\bf 1}_2+(-i\gamma^5)\right]/2 =\textrm{diag} (0,0,1,1)$.
The mode expansion of the scalar $\phi(x,y)$ 
is similar to
Eq.~(\ref{psir}).
The mode functions and the above trigonometric functions are
periodic for $y\to y +2L$.
From the properties of
orthogonality and completeness,
quadratic terms in the action integral 
(\ref{actiony}) are
\bea
 && {1\over 2}\int_{-L}^L dy\,
   \left[
    \bar{\psi} (i\gamma^\mu \partial_\mu 
     +i\gamma^5 \partial_5)\psi 
    + {1\over 2}(\partial_\mu \phi)^2 
     -{1\over 2} (\partial_5 \phi)^2 \right]
\nonumber
\\  
 &\!\!\! = \!\!\!&
   \bar{\chi}_L (i\gamma^\mu \partial_\mu) \chi_L
  +\sum_{n=1}^\infty
    \left[
 \bar{\psi}_n (i\gamma^\mu \partial_\mu -m_n)\psi_n 
      + {1\over 2}(\partial_\mu \phi_n)^2 
   +{1\over 2} (m_n \phi_n)^2 \right].
    \label{action2kk}
\eea
where $\psi_n$ is composed of $\psi_{Ln}$ and $\psi_{Rn}$  
as in the left- and right-handed projections for $\psi$.
The equation (\ref{action2kk})
is diagonal with respect to
Kaluza-Klein modes.
The Yukawa interaction is written in terms of 
four-dimensional modes as
\bea
    {1\over 2} \int_{-L}^L dy \,
     (-g) \bar{\psi}\psi \phi
   &\!\!\!=\!\!\!&
    \sum_{n=1}^\infty {(-g)\over \sqrt{L}}
  \left[ 
   \bar{\chi}_L \psi_{Rn}\phi_n 
   +
   \bar{\psi}_{Rn} \chi_L \phi_n
     \right]
\nonumber
\\
   &&+\sum_{n=1}^\infty \sum_{\ell=1}^\infty 
     \sum_{k=1}^\infty
   {(-g)\over \sqrt{2L}}
  \, \bar{\psi}_{Ln}\psi_{R\ell} \phi_{k}
   \left\{
     \delta_{k+n,\ell} +\delta_{\ell+n,k} 
     -\delta_{k+\ell,n}\right\} 
\nonumber
\\
  && +\sum_{n=1}^\infty \sum_{\ell=1}^\infty 
     \sum_{k=1}^\infty
   {(-g)\over \sqrt{2L}}
  \, \bar{\psi}_{Rn}\psi_{L\ell}\phi_{k}
    \left\{ 
     \delta_{n+\ell,k} +\delta_{k+\ell,n}
       -\delta_{n+k,\ell}\right\}
   .
\eea
The Yukawa interaction involves
the exchange between Kaluza-Klein modes.
Originally the kinetic terms are set only in the bulk.
In the next section, it will be shown
that divergence corresponding to 
brane kinetic terms is radiatively generated.

\section{Method of loop calculations \label{sec:method}}

As in four dimensions, 
wave function renormalization is expected to
appear for the fermion at the one-loop level.
The appearance of divergence on the branes 
in a similar setup is shown in Ref.~\cite{%
Georgi:2000ks}.
For the calculation of quantum loop corrections,
the steps we perform contain
the Kaluza-Klein mode expansion,
the summation of diagrams, 
the replacement of  fractions with 
a formula for the Gamma function
and the representation with
the Poisson summation for the summation of 
the Kaluza-Klein modes.
In this 
appendix, 
we give explicit equations at each step
in the method to calculate
self energy for the action integral (\ref{actiony}).
The method will be used further in 
Section~\ref{sec:subtract}.

\begin{figure}[htb]
\begin{center}
\includegraphics[width=5cm]{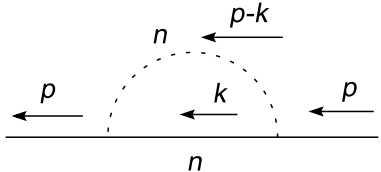}
\end{center}
\caption{Self-energy diagram for 
the left-handed zero mode.
\label{fig:chi}}
\end{figure}

The first diagram is the 
self-energy for the left-handed fermion zero mode.
The external lines are the left-handed zero mode $\chi$
and the internal lines are $\phi_n$ and $\psi_{Rn}$.
The internal lines are summed with respect to all $n$. 
The diagram is shown in Fig.~\ref{fig:chi}.
The self-energy up to the propagators
for the external lines is
\bea
   {g^2\over L} \sum_{n=1}^\infty
  \int {d^4 k\over (2\pi)^4}
   {k\!\!\!/ +m_n\over k^2 -m_n^2}
  P_L
  {1\over (p-k)^2 -m_n^2} 
  = {g^2\over L}
       \sum_{n=1}^\infty
          \int_0^1 dx \int {d^4\ell\over (2\pi)^4}
            {x p\!\!\!/ +m_n \over 
             \left[\ell^2 -\Delta_n\right]^2}
     P_L ,
     \label{eq1}
\eea
with a Feynman parameter $x$.
Here
$\ell = k-xp$ and
$\Delta_n = -x(1-x) p^2 +m_n^2$.
We will focus on $p\!\!\!/$ terms,
i.e., the terms for the wave function renormalization,
neglecting $m_n$ in the numerator.
Then the integrand is 
symmetric under $n\leftrightarrow -n$.
The $p\!\!\!/$ term in Eq.~(\ref{eq1}) is rewritten as
\bea
  {g^2\over 2L} 
       \sum_{n=-\infty}^\infty
          \int_0^1 dx \int {d^4\ell\over (2\pi)^4}
            {x (ip\!\!\!/)  \over 
             \left[\ell_E^2 +\Delta_n\right]^2}
     P_L 
  -{g^2\over 2L} 
          \int_0^1 dx \int {d^4\ell_E\over (2\pi)^4}
            {x (ip\!\!\!/)  \over 
             \left[\ell_E^2 +\Delta_0\right]^2}
     P_L ,
      \label{nand0}
\eea     
with the Wick rotation $\ell^0 =i\ell_E^0$ and
$\ell^i =\ell^i_E$.
With use of symmetry under $n\leftrightarrow -n$,
the sum in the first term has been taken for
$-\infty \leq n \leq \infty$ which is more 
convenient 
than treating the sum for $1\leq n\leq \infty$.
For this infinite sum,  a formula for the Poisson summation will be 
employed.
The second term in Eq.~(\ref{nand0}) is independent of
internal mode $n$.

Let us calculate the sum part $\sum_{n=-\infty}^\infty$.
The first step is to rewrite the fraction using the 
integral representation of the Gamma function as
\bea
 && {1\over \left[\ell_E^2 +\Delta_n\right]^2}
  ={1\over \Gamma(2)}\int_0^\infty dt \, t
    e^{-(\ell_E^2 +\Delta_n)t}
  =\int_0^\infty dt\, t
    e^{-(\ell_E^2 -x(1-x)p^2 
      +{n^2\pi^2 \over L^2})t} .
\eea
The second step is to change the mode index with the Poisson
summation,
\bea      
  && \sum_{n=-\infty}^\infty
   e^{-n^2 {\pi^2 t\over L^2}}
   ={L\over \sqrt{\pi t}}
     \sum_{n_p=-\infty}^\infty
       e^{-n_p^2 {L^2 \over t}} .
\eea
Then the sum part, the first term in Eq.~(\ref{nand0}),
is written as
\bea
    {g^2
  \over 2^{5} \pi^{5\over 2}}       
          \int_0^1 dx \,
            x (ip\!\!\!/)  
         \sum_{n_p=-\infty}^\infty \int_0^\infty dt 
     \,t^{-{3\over 2}}
           e^{x(1-x)p^2 t-{n_p^2 L^2 \over t}}
     \,P_L,          
\eea
where the Gaussian integral for $\ell_E$
have been employed. 
The next step is to evaluate each part for $n_p=0$
and $n_p\neq 0$.
The $n_p=0$ part includes
divergence for $t\to 0$. 
The divergence of the $t$ integral is evaluated as
\bea
   \int_0^\infty dt \, t^{-{3\over 2}}
     (1+ x(1-x)p^2 t+\cdots)
   =\left[ -2 t^{-{1\over 2}} +\cdots
  \right]_{t=\Lambda^{-2}}^{t=\infty}
   =2\Lambda + \textrm{finite} ,
\eea
at the cutoff regularization.
This momentum integral depends on
the regularization scheme.
The scheme-independence of the renormalized quantity
will be shown in Section~\ref{sec:momentum}.
The divergence of the $n_p=0$ part is
\bea
  {g^2\over 2^5 \pi^{5\over 2}}
      \int_0^1 dx \, x(ip\!\!\!/)
     \cdot  2\Lambda \, P_L 
 = {g^2  \over 2L(4\pi)^2}
   {\Lambda L\over \sqrt{\pi}}
    (ip\!\!\!/) P_L .
  \label{div1}
\eea
From the mass dimensions for the fields and coupling
$[\psi]=[\textrm{mass}]^2$, 
$[\phi]=[\textrm{mass}]^{3/2}$,
$[g]=[\textrm{mass}]^{-{1/2}}$,
the bulk divergence for the wave function is expected to have
the coefficient $g^2 \Lambda$
where $[g^2 \Lambda]=[\textrm{mass}]^0$ 
and it is consistent with
Eq.~(\ref{div1}).
The $n_p\neq 0$ part is
\bea
 {g^2 \over 2^3 \pi^{5\over 2}}
    \int_0^1 dx \, x(ip\!\!\!/)
    \sum_{n_p=1}^\infty
       {\sqrt{\pi}\over 2n_p L}
      e^{-2n_p L{\sqrt{-x(1-x)p^2}}}
        P_L ,
        \label{npn1}
\eea       
where the integral for $t$ has been performed 
with help of expressions for Bessel function $K_\nu$.
If a function $F(x)$ is symmetric 
under the exchange $x\leftrightarrow 1-x$, $F(x)=F(1-x)$,
then $\int_0^1 dx \, x F(x)=\int_0^1 dx \, (1-x) F(x)
 =(1/2)\int_0^1 dx \, F(x)$.
The equation (\ref{npn1}) becomes
\bea
  - {g^2 \over 2L (4\pi)^2}
       \int_0^1 dx \, (ip\!\!\!/)
      \log \left[
          1- e^{-2 L\sqrt{-x(1-x)p^2}} 
    \right] \cdot P_L ,
\eea
where the sum for $n_p$ has been performed
with the Taylor series expansion 
$\log (1-X) =-\sum_{s=1}^\infty {X^s /s}$.

Now we move on to the second line in Eq.~(\ref{nand0}),
\bea
  (\Delta_0 ~\textrm{part})
  =-{g^2\over 2L} \int_0^1 dx 
    \int {d^4\ell_E \over (2\pi)^4}
    {x(ip\!\!\!/)\over \left[
      \ell_E^2 +\Delta_0\right]^2}
      P_L ,
      \label{delta0}
\eea
where $\Delta_0 = -x(1-x)p^2$.
This is a usual four-dimensional integral.
For the cutoff regularization,
Eq.~(\ref{delta0}) is given  by
\bea
   -{g^2 \over 2L (4\pi)^2}
  (ip\!\!\!/)
  \left(\log(\Lambda L)
   -{1\over 2} \log(-p^2 L^2) +1 \right) \cdot
    P_L .
\eea
This divergence is expected from the dimensional counting
$[g^2]=[\textrm{mass}]^{-1}$ and 
$[1/L]=[\textrm{mass}]$. 
For the self-energy with the field $\chi$ in the external lines,
we have found linear divergence in the summation part 
$\sum_{n=-\infty}^\infty$ and logarithmic divergence in the
non-summation part.

The self-energy with the field $\psi_L$
in the external lines is analogous.
Here there are various Kaluza-Klein modes 
to contribute.
The internal lines are composed of $\phi$ and $\psi_R$.
The Kaluza-Klein modes for the external and internal lines
are labeled as shown in Figure~\ref{fig:kkl}.
%
%
The momentum flow is the same as in Figure~\ref{fig:chi}.
With the same procedure as in obtaining
Eq.~(\ref{nand0}) from Eq.~(\ref{eq1}) for the diagram in
Figure~\ref{fig:chi}, the part $p\!\!\!/$ in
the sum of the diagrams (a), (b) and (c) in Figure~\ref{fig:kkl}
is obtained as
\bea
   ( \textrm{(a)} +\textrm{(b)} +
    \textrm{(c)} )|_{p\!\!\!/}
  &\!\!\!=\!\!\!& {g^2\over 2L}
       \sum_{s=-\infty}^\infty
        \int_0^1 dx
       \int {d^4 \ell\over (2\pi)^4} \,
       {x p\!\!\!/ \over 
     \left[\ell^2 -\Delta_{s+}\right]^2}
        P_L
\nonumber
\\
  &&  
   - {g^2\over 2L}
        \int_0^1 dx
       \int {d^4 \ell\over (2\pi)^4} \,
       {x p\!\!\!/ \over 
     \left[\ell^2 -\Delta_{0+}\right]^2}
        P_L   
\nonumber
\\        
  && - {g^2\over 2L}
        \int_0^1 dx
       \int {d^4 \ell\over (2\pi)^4} \,
       {x p\!\!\!/ \over 
     \left[\ell^2 -\Delta_{n+}\right]^2}
        P_L .
    \label{sumabc}
\eea
Here
$\Delta_{s+}
= -x(1-x) p^2 + 
(s+(1-x)n)^2 \pi^2/L^2
+x(1-x)n^2 \pi^2/L^2$.
The other $\Delta_{0+}$ and $\Delta_{n+}$ are
obtained from $\Delta_s+$ as
$\Delta_{0+} =\Delta_{s+}|_{s=0}$
and $\Delta_{n+} =\Delta_{s+}|_{s=n}$.
The divergent part is given by
\bea
  ((\textrm{a})+(\textrm{b})+(\textrm{c}))|_{\textrm{%
    \scriptsize div}}
  &\!\!\!=\!\!\!&
  {g^2 \over 2L(4\pi)^2}
  (ip\!\!\!/)
  \left[ {\Lambda L\over \sqrt{\pi}}
   -\log (\Lambda^2 L^2)\right] \cdot P_L .
   \label{divabc}
\eea   
For the diagrams (d), (e), (f) and (g) in Figure~\ref{fig:kkl},
the external lines have $n$ and $n+2s$.
The modes $n$ and $s$ in 
the internal lines are fixed for given external lines.
These correspond to the non-sum part in Eq.~(\ref{nand0}).
These diagrams satisfy
$\textrm{(g)}=\textrm{(d)}$
and $\textrm{(f)}=\textrm{(e)}$.
The divergent part is
\bea
   (\textrm{d})|_{\textrm{%
     \scriptsize div}}
   =(\textrm{g})|_{\textrm{%
     \scriptsize div}}
   &\!\!\!=\!\!\!&
     -{g^2\over 4L(4\pi)^2} 
       (ip\!\!\!/) \log (\Lambda^2 L^2) \cdot P_L 
\nonumber
\\
  &\!\!\!=\!\!\!&
     (\textrm{f})|_{\textrm{%
     \scriptsize div}} 
   =(\textrm{e})|_{\textrm{%
     \scriptsize div}} .
     \label{divdefg}
\eea
The self-energy with the fields $\chi$ and $\psi_L$
in the external lines is similarly calculated.

The other diagrams are the ones that have 
the right-handed field $\psi_R$ in the 
external lines.
Similarly to diagrams with $\psi_L$ in an external line,
the Kaluza-Klein modes are labeled.
For (a), (b), $\cdots$, (g),
there are 
the corresponding diagrams
(a$'$), (b$'$), $\cdots$, (g$'$).
The amplitudes (a$'$), (b$'$), (c$'$), (d$'$) and (g$'$)
are given by (a), (b), (c), (d) and (g) with
$P_L$ replaced by $P_R$.
The amplitudes (e$'$) and (f$'$) are 
given by (e) and (f) with $P_L$ replaced by $P_R$
and with the change of the overall sign.
In addition to these diagrams, there is a diagram
with $\psi_R$ in the external lines and with
$\chi$ and $\phi$ in the internal lines.
This contributes to the divergent part as
\bea
   {g^2 \over 2L(4\pi)^2} (ip\!\!\!/) \log (\Lambda^2 L^2) \cdot
     P_R .
     \label{rchi}
\eea
The equation (\ref{rchi}) and the last term in Eq.~(\ref{divabc}) with the replacement $P_L$ by $P_R$
cancel each other.
The contributions (d$'$) and (e$'$) cancel each other
and (f$'$) and (g$'$) cancel each other.
In summary for the right-handed external lines,
the remaining divergence is
\bea
   {g^2\over 2L (4\pi)^2}
      (ip\!\!\!/) {\Lambda L\over \sqrt{\pi}} \cdot P_R .
      \label{divpsir}
\eea
The logarithmic divergence for the 
external $\psi_R$
is canceled out.

The final step in the loop calculation is 
to identify Lagrangian terms corresponding to 
the divergence.
For all the divergences (\ref{divabc}),
(\ref{divdefg}) and (\ref{divpsir}),
the Lagrangian terms are given by
\bea
 &&
 {(-i)g^2 \over 2L (4\pi)^2} {\Lambda L\over \sqrt{\pi}}
  \left( \bar{\chi}_L (ip\!\!\!/) \chi_L
   +\sum_n \bar{\psi}_n (ip\!\!\!/) \psi_n\right)
\nonumber
\\
  && -{(-i)g^2 \over 2L (4\pi)^2}
  \log (\Lambda L)
  \bigg\{
   \bar{\chi}_L (ip\!\!\!/) \chi_L
   +2 \sum_n \bar{\psi}_{n,L} (ip\!\!\!/)
  \psi_{n,L}
\nonumber
\\
  && +2 \sum_n \sum_s 
 \left(\bar{\psi}_{n,L} (ip\!\!\!/) \psi_{n+2s,L}
   +\bar{\psi}_{n+2s,L} (ip\!\!\!/) \psi_{n,L}\right)
\nonumber
\\
   && +\sqrt{2} \sum_n
   \left(\bar{\chi}_L (ip\!\!\!/)\psi_{2n,L}
 +\bar{\psi}_{2n,L} (ip\!\!\!/) \chi_L \right)
  \bigg\} .
\eea   
From this equation, the Lagrangian terms 
with four-derivative
are written 
in terms of the five-dimensional field
as
\bea
 {g^2 \over 2L(4\pi)^2}
   \cdot {1\over 2}\lim_{\epsilon\to 0}
     \int_{-L+\epsilon}^{L+\epsilon}
     dy 
     \left[
      {\Lambda L\over \sqrt{\pi}} \,
        \bar{\psi}(i\partial\!\!\!/)
        \psi
        -\log (\Lambda L) \,
          \bar{\psi}_L
          (i\partial\!\!\!/)
          \psi_L
          \left(\delta(y) +\delta(y-L)\right)\right]
          , \label{simpletotal}
\eea
where $\lim_{\epsilon\to 0}
 \int_{-L+\epsilon}^{L+\epsilon}
 dy \, G(y) \left[\delta(y)+\delta(y-L)\right]
 =G(0) +G(L)$.
The divergence in the brane terms for one-loop
wave function occurs 
in the equal size at $y=0$ and at $y=L$,
in agreement with Ref.~\cite{%
Georgi:2000ks}.
For the boundary condition (\ref{psiLR}), 
the right-handed component
$\psi_R$ does not have the brane term.
In the equation~(\ref{simpletotal}),
the brane divergence includes 
non-diagonal components with
respect to Kaluza-Klein modes and 
the bulk divergence is diagonal.

\section{Mode functions \label{app:mfg}}

According to the approach given in
Ref.~\cite{Dvali:2001gm}, 
we derive $f_n$ and $g_n$.
The fermion mode functions appearing in the context,
$f_n$ and $g_n$, are $f_{n\,\textrm{\scriptsize III}}$
and $g_{n\, \textrm{\scriptsize III}}$
given in this appendix.

\begin{figure}[htb]
\hspace{2.5cm}
\includegraphics{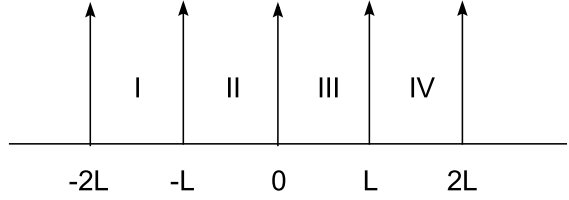}
\caption{Regions. Allows depict $\delta$ functions\label{fig:reg}}
\end{figure}

From the equations of motion (\ref{eom1}) and
(\ref{eom2}) and Eq.~(\ref{modee}),
the mode functions obey
\bea
    \partial_y f_n &\!\!\!=\!\!\!&
    -m_n g_n ,
\\
   \partial_y g_n &\!\!\!=\!\!\!&
  m_n \left(1+ a\, \left(\delta(y) +\delta(y-L)\right)\right) f_n ,
\eea
for $n\neq 1$.  
From these equations,
the second-order differential equation 
for $f_n$ is obtained as
\bea
  \left(\partial_y^2 + m_n^2
   +a\, m_n^2 \left( \delta(y) + \delta(y-L)\right) \right)
    f_n
  =0 .
    \label{second}
\eea 
Unless confusion arises,
we will omit the subscript for 
the Kaluza-Klein mode number $n$ for $f_n$.
The equation (\ref{second})
includes two delta functions.
It is convenient to write the function by defining separate regions
in a period $2L$.
The regions can be classified as shown in Figure~\ref{fig:reg}.
In each region, the function $f$ where 
is written as
\bea
   f_{\textrm{\scriptsize I}}
     (y) &\!\!\!=\!\!\!&
       A e^{imy} + Be^{-imy} ,
\\
  f_{\textrm{\scriptsize II}} 
     (y) &\!\!\!=\!\!\!&
        C e^{imy} + D e^{-imy} ,
\\
   f_{\textrm{\scriptsize III}}
     (y) &\!\!\!=\!\!\!&
        E e^{imy} + Fe^{-imy} ,
\\
    f_{\textrm{\scriptsize IV}}
      (y) &\!\!\!=\!\!\!&
         Ge^{imy} + He^{-imy} ,
\eea
where $A,B,C,D,E,F,G,H$ are 
determined by 8 conditions.
They are periodicity between the regions
I and III and between the regions II and IV, 
continuity at $y=-L,0,L$,
matching of the first derivative at $y=-L,0,L$ and 
the normalization.
Among the above 9 conditions, 
one condition is automatically satisfied if the other conditions
are satisfied.

Now we fix these constants.
The periodicity requires
\bea
   &&
    A e^{im (y-2L)}
      +Be^{-m(y-2L)}
       =E e^{imy}
       +Fe^{-imy} ,
\\
  && 
   C e^{im(y-2L)}
    +D e^{-m(y-2L)}
    =G e^{imy}
     +H e^{-imy} .
\eea
The continuity requires
\bea
 &&A e^{-imL} + Be^{imL} = C e^{-imL} + D e^{imL} ,
  \label{cont-l}
\\
  &&C+D = E+F ,
   \label{cont0}
\\
  && E e^{imL} + F e^{-imL}
   = G e^{imL} + H e^{-imL} ,
   \label{contl}
\eea
at $y=-L,0,L$, respectively.
The matching of first derivative requires
\bea
  && \left[\partial_y f_n\right]_{\textrm{\scriptsize I}}^{\textrm{%
   \scriptsize II}}
     +a \, m_n^2 f_{n\textrm{%
  \scriptsize I}} |_{y=-L}
   = 0 ,
     \label{matching-l}
\\
   && \left[\partial_y f_n\right]_{\textrm{%
   \scriptsize II}}^{\textrm{\scriptsize III}}
    + a m_n^2 f_{n \textrm{\scriptsize II}} |_{y=0} =0 ,
   \label{matching0}
\\
  &&
   \left[\partial_y f_n\right]_{\textrm{%
    \scriptsize III}}^{\textrm{\scriptsize IV}}
     +a m_n^2 f_{n\textrm{\scriptsize III}}|_{y=L}
     = 0 ,
       \label{matchingl}
\eea
at $y=-L,0,L$, respectively. 
Here $f_{n\textrm{\scriptsize I}}$ in
the second term in Eq.~(\ref{matching-l})
can be replaced by
$f_{n\textrm{\scriptsize II}}$ 
because of the continuity
(\ref{cont-l}).
The equations (\ref{matching0}) and (\ref{matchingl})
are understood similarly.
In summary, the conditions up to normalization
are
\bea
  &&
  A e^{im(y-2L)} + Be^{-im (y-2L)}
   = E e^{imy} + F e^{-imy} ,
   \qquad (\textrm{for}~ 0_+ \leq y\leq L_-) ,
    \label{cd1}
\\
  &&
   C e^{im(y-2L)}
  + D e^{-im(y-2L)}
    = G e^{imy} + H e^{-imy} ,
  \qquad (\textrm{for} ~ L_+ \leq y \leq 2L_-) , 
   \label{cd2} 
\\
   &&
   A e^{-imL} + Be^{imL} 
  = C e^{-imL} + De^{imL} ,
    \label{cd3}
\\
  && C+ D =E+F ,
  \label{cd4}
\\
  && E e^{imL} + F e^{-imL}
    = G e^{imL} + H e^{-imL} ,
    \label{cd5}
\\
   &&
   C e^{-imL} - D e^{imL}
    -A (1 + iam) e^{-imL} + B(1-iam) e^{imL} = 0, 
    \label{cd6}
\\
  &&
   E-F -C (1+iam) + D (1-iam) = 0,
   \label{cd7}
\\
   &&
  Ge^{imL} -H e^{-imL}
  -E(1+iam) e^{imL}
   +F(1-iam) e^{-imL} = 0,
   \label{cd8}
\eea 
where $0_+ = \lim_{\epsilon\to 0} 0+\epsilon$ and
$L_\mp =\lim_{\epsilon \to 0} L\mp \epsilon$.
Using the first three equations~(\ref{cd1}), 
(\ref{cd2}) and (\ref{cd3}),
the fifth condition (\ref{cd5}) is automatically satisfied.
Hence, the above 7 conditions and the normalization fix 
the 8 constants $A,B,\cdots, H$.
Up to the normalization,
the solution of the mode function $f$ is
given by
\bea
  f_{\textrm{\scriptsize I}} (y)
    &\!\!\!=\!\!\!&
    A (e^{i my} + w^3 e^{-i my}) ,
\\
  f_{\textrm{\scriptsize II}} (y)
     &\!\!\!=\!\!\!&
       A (w e^{i my} + w^2 e^{-i my}) ,
\\
  f_{\textrm{\scriptsize III}} (y)
    &\!\!\!=\!\!\!&
      A (w^2 e^{i my} + w e^{-i my}) ,
\\
  f_{\textrm{\scriptsize IV}} (y)
   &\!\!\!=\!\!\!&
      A (w^3 e^{i my} +e^{-i my}) ,
\eea
with the mass quantization condition
\bea
   -\tan mL = {am\over 1-\left({am\over 2}\right)^2} ,
   \label{massquant}
\eea
which is also written as
$e^{-2imL} = (2+iam)^2/(2-iam)^2$.
Here
$w_n \equiv (2+i am_n)/(2 - i am_n)$.

The orthogonality of the mode function can be found as follows.
The mode function satisfies
the second-order differential equation
(\ref{second}).
From this equation, the following equation is 
derived:
\bea
 0 &\!\!\!=\!\!\!& {1\over 2} \lim_{\epsilon\to 0}
   \int_{-L+\epsilon}^{L+\epsilon} dy \, \bigg\{
    f_m \left(\partial_y^2 +m_n^2 
     +am_n^2 \left(\delta(y) + \delta(y-L)\right)\right) f_n
\nonumber
\\
  && -f_n \left(\partial_y^2 +m_m^2 
     +am_m^2 \left(\delta(y) + \delta(y-L)\right) \right) f_m
     \bigg\} .
   \label{wronsk}
\eea
It is found that the mode function satisfies
\bea
  \lim_{\epsilon\to 0} (f_m \partial_y f_n )_{\textrm{%
    \scriptsize IV}, L+\epsilon}
  &\!\!\!=\!\!\!& A_m A_n
     (w_m +1) ^{-im_m L} (w_n-1) im_n e^{-im_nL}
\nonumber
\\
  &\!\!\!=\!\!\!&
   \lim_{\epsilon\to 0}
  (f_m\partial_y f_n)_{\textrm{%
   \scriptsize II}, -L+\epsilon} ,
\eea
and 
\bea
    \int_{-L}^0 dy \, f_{\textrm{\scriptsize II} n} (y)
      f_{\textrm{\scriptsize II} m}(y)
      =
        \int_L^0 d(-y)
         \,
         f_{\textrm{\scriptsize II} n}(-y)
         f_{\textrm{\scriptsize II} m}(-y)
 =
     \int_0^L dy \, 
       f_{\textrm{\scriptsize III} n} (y)
        f_{\textrm{\scriptsize III} m} (y) .
\eea
Using these equations,
Eq.~(\ref{wronsk}) becomes
\bea 
  0 = (m_n^2 -m_m^2)
   \left\{
  \int_{0}^{L}
   dy \, f_n (y) f_m (y)
  + {a\over 2} \left( f_n(0) f_m(0) + f_n (L) f_m (L)\right) 
  \right\} ,
\eea
where
$\lim_{\epsilon \to 0} \int_L^{L+\epsilon} dy \,
  (\textrm{smooth function}) =0$.
Therefore the orthogonality is 
given in Eq.~(\ref{ortf}).
Here $A$ is normalized as
\bea
   2A^2 w^3 \left( L + {4a \over 4+ a^2 m^2}\right)
     =1 .
\eea
The completeness corresponds to
\bea
   \sum_m f_m(y) f_m(y')
  = 2 \delta (y-y')
    -  a \, \left(\delta(y) +\delta(y-L)\right)
    \sum_m f_m (y) f_m (y') .
\eea
The mode function $g(y)$ is derived from 
the relation $g=-(1/m) \partial_y f$.
They are
\bea
  g_{\textrm{\scriptsize I}} (y)
    &\!\!\!=\!\!\!&
     -A i (e^{i my} -w^3 e^{-i my}) ,
\\
  g_{\textrm{\scriptsize II}} (y) 
    &\!\!\!=\!\!\!&
       -A i (w e^{i my} -w^2 e^{-i my}) ,
\\
  g_{\textrm{\scriptsize III}} (y)
    &\!\!\!=\!\!\!&
       -A i (w^2 e^{i my} -w e^{-i my}) ,
\\
  g_{\textrm{\scriptsize IV}} (y) 
    &\!\!\!=\!\!\!&
  -A i (w^3 e^{i my} - e^{-i my}) .
\eea
The function $g$ has the orthogonality
given in Eq.~(\ref{ortg}).
Here $+$ in $\pm$
has been adopted as $w_n = e^{-i my}$ 
in $\pm e^{-i my}$. 
In the limit $a=0$,
these mode functions reduce to the mode functions 
in Eqs.~(\ref{psil}) and (\ref{psir}).

\end{appendix}

\newpage



\end{document}